\documentstyle[11pt,IAUS212,twoside,epsf]{article}

\def\edcomment#1{\iffalse\marginpar{\raggedright\sl#1\/}\else\relax\fi}
\reversemarginpar

\newcommand{\mdot}{$\dot{M}$}
\newcommand{\vinf}{$v_{\infty}$}

\begin{document}
\vspace*{1cm}
\title{Mass-loss predictions and stellar masses of early-type stars}
 \author{A. de Koter}
\affil{Astronomical Institute ``Anton Pannekoek'', University of Amsterdam,
       Kruislaan 403, NL-1098 SJ Amsterdam, The Netherlands}
\author{J.S. Vink}
\affil{Imperial College, Blackett Laboratory, Prince Consort Road, London,
       SW7 2BZ, U.K.}

\begin{abstract}
  We show that the stellar masses implied by our predictions of the
  wind properties of massive stars are in agreement with masses derived
  from evolution theory and from direct measurements using spectroscopic
  binaries, contrary to previous attempts to derive masses from wind
  theory.
\end{abstract}

\section{Introduction}

  The stellar winds of early-type stars are thought to be driven by
  radiation pressure on spectral lines. Predictions of wind properties 
  are usually tested by comparing them to \mdot\ \& \vinf\ values 
  derived for a set of about 30 well studied Galactic and Magellanic 
  Cloud O and early-B stars (Puls et al. 1996). Although it has proven
  challenging enough to match predictions with observations, these
  stars provide only a limited test.
  To better constrain wind theory one should confront a wider range
  of predictions with observations.
  Meaningful new tests include extending the comparisons of \mdot\ and 
  \vinf\ to:
  {\em   i)} extremely luminous Of and WN5h-6h stars;
  {\em  ii)} stars at (both sides of) the bistability jump, an abrupt 
             discontinuity in \vinf\ found to occur at 
             spectral type B1, and
  {\em iii)} Luminous Blue Variables.
  We have applied our predictions of mass loss to all the above cases
  with excellent results (de Koter et al. 1997; Vink et al. 1999,
  2000, 2002). 

  As the properties of stellar winds depend also on stellar mass, an 
  alternative test would be to compare masses derived from line-driven
  wind theory with those of independent methods. This is a relevant
  issue in view of current problems with masses based on
  mass-loss rates and terminal wind velocities.

\section{Stellar masses derived from wind theory}

\begin{figure}
\begin{center}
\epsfxsize=7.2cm
\epsfbox{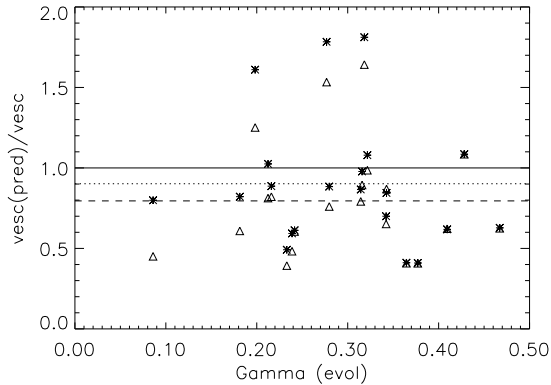}
\caption{Consistency check for O stars between predicted surface escape velocity 
          and $v_{\rm esc}$ implied by the stellar parameters. 
          See text for a discussion.
          }           
\end{center}
\end{figure}

  Although some wind properties (of O stars) are almost mass independent, 
  such as the modified wind momentum $\dot{M} v_{\infty} \sqrt{R}$, the mass 
  loss itself behaves as $\dot{M} \propto M_{\ast}^{-1.3}$ (Vink et al. 2000). 
  So, in principle, the mass may be derived when all relevant stellar 
  properties -- including \mdot\ \& \vinf\ -- are known. So far, attempts 
  to derive $M$ from wind theory yielded systematically lower masses
  (up to tens of percent) than those derived from evolutionairy tracks,
  but appeared to be in agreement with spectroscopic mass determinations
  (e.g. Herrero et al. 2001).
  
  Using the latest generation of model atmospheres (see Herrero this
  proceedings) the spectroscopic masses, however, now appear to be in
  much better
  agreement with $M_{\rm evol}$. Also, direct mass determinations using
  spectroscopic binaries (e.g. Massey et al. 2002) seem to favor 
  evolutionairy masses. This would leave the wind masses as the exception.
  However, here we present evidence that also our mass-loss predictions favor 
  $M_{\rm evol}$. 
  The figure shows a consistency check between predicted
  surface escape velocity $v_{\rm esc} \propto [M_{\ast}(1-\Gamma)]^{1/2}$ 
  and $v_{\rm esc}$ implied by the stellar parameters. $\Gamma$ is the 
  Eddington factor for electron scattering. The prediction involves the 
  observed mass loss and terminal wind velocity, i.e. 
   $v_{\rm esc}({\rm pred}) = f(\dot{M},v_{\infty},L_{\ast},
                              T_{\rm eff}, M_{\ast})$. 
  The case for the evolutionairy masses is shown using asterisk symbols; 
  ``old'' spectroscopic masses (in agreement with previous attempts to 
  derive masses from wind theory) have triangle symbols. The mean ratio 
  for $M_{\rm evol}$ (dotted line) is $0.90 \pm 0.16$; for $M_{\rm spec}$ 
  (dashed line) it is $0.79 \pm 0.12$.
  
  This shows that within the error our predicted wind properties are in
  agreement with evolutionairy masses, bringing also the masses from
  wind theory in accord with those derived from all other
  mass determination methods.

\end{document}